\documentclass[twocolumn,aps,prl,preprintnumbers]{revtex4}
\usepackage{}
\usepackage{bbm}
\usepackage[latin9]{inputenc}
\usepackage{amsmath}
\usepackage{amssymb}
\usepackage{graphicx}
\usepackage{mathrsfs}
\usepackage{amsfonts}
\usepackage{amsthm}
\usepackage{color}
\usepackage{txfonts}
\usepackage{lipsum}
\usepackage{gensymb}
\usepackage[colorlinks=true,citecolor=blue,linkcolor=blue,urlcolor=blue,anchorcolor=blue]{hyperref}%
\hypersetup{colorlinks=true,citecolor=blue,linkcolor=blue,urlcolor=blue}
\providecommand{\U}[1]{\protect\rule{.1in}{.1in}}
\makeatletter

\newcommand{\Rmnum}[1]{\expandafter\@slowromancap\romannumeral #1@}
\makeatother
\makeatletter
\@ifundefined{textcolor}{}
{
\definecolor{BLACK}{gray}{0}
\definecolor{WHITE}{gray}{1}
\definecolor{RED}{rgb}{1,0,0}
\definecolor{GREEN}{rgb}{0,1,0}
\definecolor{BLUE}{rgb}{0,0,1}
\definecolor{CYAN}{cmyk}{1,0,0,0}
\definecolor{MAGENTA}{cmyk}{0,1,0,0}
\definecolor{YELLOW}{cmyk}{0,0,1,0}
}
\makeatother
\begin{document}
\title{Robust edge states in magnetic domain-wall racetrack}
\author{Z.-X. Li}
\author{Zhenyu Wang}
\author{Yunshan Cao}
\author{H. W. Zhang}
\author{Peng Yan}
\email[Corresponding author: ]{yan@uestc.edu.cn}
\affiliation{School of Electronic Science and Engineering and State Key Laboratory of Electronic Thin Films and Integrated Devices, University of Electronic Science and Technology of China, Chengdu 610054, China}
\begin{abstract}
Controllable artificial pinning is indispensable in numerous domain-wall (DW) devices, such as memory, sensor, logic gate, and neuromorphic computing hardware. The high-accuracy determination of the effective spring constant of the pinning potential, however, remains challenging, because the extrinsic pinning is often mixed up with intrinsic ones caused by materials defects and randomness. Here, we study the collective dynamics of interacting DWs in a racetrack with pinning sites of alternate distances. By mapping the governing equations of DW motion to the Su-Schrieffer-Heeger model and evaluating the quantized Zak phase, we predict two topologically distinct phases in the racetrack. Robust edge state emerges at either one or both ends depending on the parity of the DW number and the ratio of alternating intersite lengths. We show that the in-gap DW oscillation frequency has a fixed value which depends only on the geometrical shape of the pinning notch, and is insensitive to device imperfections and inhomogeneities. We propose to accurately quantify the spring coefficient that equals the square of the robust DW frequency multiplied by its constant mass. Our findings suggest as well that the DW racetrack is an ideal platform to study the topological phase transition.
\end{abstract}

\maketitle
\emph{Introduction.}---Magnetic domain walls (DWs) \cite{AtkinsonNM2003,CatalanRMP2012,KoyamaNM2011} have attracted tremendous recent attention owing to their fundamental nature and potential applications \cite{AllwoodS2005,XuNN2008,OmariPRA2014,LuoN2020,OnoAPE2008,BisigAPL2009,HeAPL2007,MartinezPRB2011,ParkinS2008,ParkinNN2015,UmmelenSR2017,PolenciucAPL2014,ShenAPL2018,ZhangAPL2014,BoriePRA2017}. To achieve a reliable control of the DW position in spintronic devices, engineered pinning sites are often introduced by artificial notches \cite{HayashiPRL2006,LepadatuPRL2009,DolocanAPL2014,YuanPRB2014,BedauPRL2008}, protrusions \cite{ZengAPL2009,LewisAPL2011,ChangAPL2012,FranchinPRB2011,PetitJAP2008}, kinks \cite{GlathePRB2012}, etc. It has been shown that the shape and strength of pinning potential can strongly affect the DW dynamics \cite{AkermanPRB2010}. For instance, a critical current must be overcome to depin the DW \cite{HeJAP2005,RavelosonaPRL2005}. The approach was adopted as well to quantify the spin-transfer torque (STT) non-adiabaticity $\beta$ \cite{LepadatuPRB2009,LepadatuPRB2010}, an important quantity for current-induced DW motion while its true value is still highly controversial \cite{LepadatuPRB2009,LepadatuPRB2010,MeierPRL2007,MoriyaNPhys2008,HeynePRL2010,ThomasN2006,BisigPRL2016}. One reason for the inconsistency reported in experiments may come from the lack of reliable determination of the pinning potential.

In principle, by measuring the oscillation frequency of a pinned DW, one can obtain the strength of the confining potential \cite{ThomasN2006}. However, the extrinsically engineered pinning is often mixed up with the intrinsic ones originating from materials defects and randomness \cite{JiangPRB2013,BadeaPRA2016}, leading to a seemingly insuperable difficulty to distinguish the two contributions. A high-accuracy determination of the profile of the pinning potential is therefore vital not only for a consistent and deep understanding of the STT physics but also for an efficient and reliable design of DW-based spintronic devices. Topology theory often bridges the gap between the global and local characteristics of a physical system. One of the most outstanding examples is the quantum Hall effect \cite{Klitzing1980}, which defines a resistance that depends only on fundamental physical constants due to the robust in-gap edge states, making possible an accurate and standardized definition of the ohm. It thus motivates us to pursue a topological method for the realization of the DW-frequency standard that is immune from the intrinsic disorder and defects.

\begin{figure}[!htbp]
\begin{centering}
\includegraphics[width=0.48\textwidth]{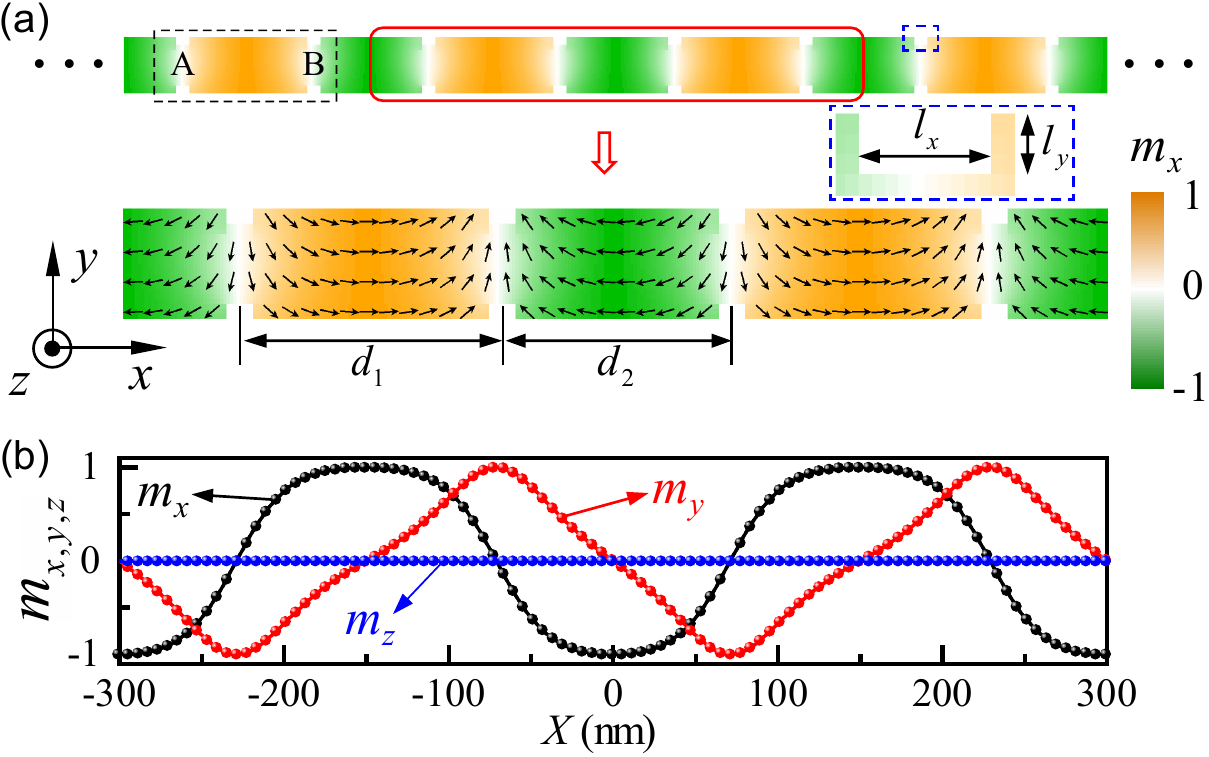}
\par\end{centering}
\caption{(a) Illustration of a DW lattice in a ferromagnetic nanostrip with width $L_{y}=60$ nm and thickness $L_{z}=5$ nm. The enlarged image below shows the micromagnetic structure of N$\acute{\text{e}}$el-type DWs pinned by periodic cuboid notches with $l_{x}=20$ nm, $l_{y}=9$ nm, and $l_{z}=5$ nm. The unit cell contains two DWs at sites A and B. $d_{1}$ and $d_{2}$ are the intracellular and intercellular distances between notches, respectively. (b) The components of normalized magnetization along the center of nanostrip ($y=30$ nm) with $d_{1}=160$ nm and $d_{2}=140$ nm.}
\label{Figure1}
\end{figure}

In this work, we theoretically investigate the collective dynamics of DWs that are locally pinned by notches with alternate distances in a magnetic nanostrip racetrack [see Fig. \ref{Figure1}(a)]. Without loss of generality, we consider the N$\acute{\text{e}}$el-type DW in the setup. The governing equation of DW motion is mapped to the Su-Schrieffer-Heeger model that allows a topological description. By evaluating the quantized Zak phase, we predict two topologically distinct phases in the racetrack. The bulk-boundary correspondence dictates a robust DW oscillation at the racetrack edges. We show that the oscillating frequency of the edge DW has a constant value which depends only on the geometrical shape of the pinning notch, and is insensitive to material imperfection and inhomogeneity. We perform full micromagnetic simulations to verify theoretical predictions with a great agreement. Our results offer the standard of the DW oscillation frequency and suggest as well that the magnetic DW racetrack is an ideal platform to study the fundamental topological phase and phase transition, in addition to its practical application in information technology.

\emph{Model.}---We start with the Landau-Lifshitz-Gilbert equation that governs the DW dynamics \cite{ZhangPRL2004,ThiavilleEPL2005}:
\begin{equation}\label{Eq1}
\frac{\partial\mathbf{m}}{\partial t}=-\gamma\mathbf{m}\times\mathbf{H}_{\text{eff}}+\alpha\mathbf{m}\times\frac{\partial\mathbf{m}}{\partial t}+\mathbf{\Gamma}_{\text{st}},
\end{equation}
where $\mathbf{m}=\mathbf{M}/M_{s}$ is the unit magnetization vector with the saturated magnetization $M_{s}$, $\gamma$ is the gyromagnetic ratio, and $\alpha$ is the Gilbert damping constant.
The effective field $\mathbf{H}_{\mathrm{eff}}$ comprises the external field, the exchange field, the magnetic anisotropic field, and the dipolar field. $\mathbf{\Gamma}_{\text{st}}$ is the torque due to the spin-transfer or spin-orbit effects. For the case of STT, $\mathbf{\Gamma}_{\text{st}}=b_{J}(\hat{J}\cdot\nabla)\mathbf{m}-\beta b_{J}\mathbf{m}\times(\hat{J}\cdot\nabla)\mathbf{m}$ with $b_{J}=JPg\mu_{B}/2|e|M_{s}$ and $\hat{J}$ being the flow direction of the spin-polarized current. Here $J$ is the charge current density, $P$ is the spin polarization, $g$ is the $g$-factor, $\mu_{B}$ is the Bohr magneton, and $e$ is the (negative) electron charge.

The collective-coordinate or $\{q,\phi\}$ method provides a simple, yet accurate description of the motion of complex DWs [see Fig. \ref{Figure1}(b)], in terms of its position $q$ and tilting angle $\phi$ \cite{WalkerJAP1974,LiPRB2004}:
\begin{equation}\label{Eq2}
\begin{aligned}
 (1+\alpha^{2})\frac{dq_{j}}{dt} =&\gamma \alpha H_{\text{pin},j}\Delta_{j}+\frac{1}{2}\gamma(N_{z}-N_{y})\Delta_{j} M_{s}\text{sin}2\phi_{j},\\
  (1+\alpha^{2})\frac{d\phi_{j}}{dt} =&\gamma H_{\text{pin},j}-\frac{1}{2}\gamma\alpha(N_{z}-N_{y})M_{s}\text{sin}2\phi_{j},
\end{aligned}
\end{equation}
where the collective coordinates $q_{j}$ and $\phi_{j}$ are the position and tilt angle of the $j$-th DW, respectively, $H_{\text{pin},j}$ includes the pinning field from both the notch and the DW-DW interaction, $N_{y}$ and $N_{z}$ are the demagnetizing factors along the $y$- and $z$-axis of the nanostrip, respectively, and $\Delta_{j}=\sqrt{2A/\left\{2K_{u}+\mu_{0}M_{s}^{2}\left[(N_{y}-N_{x})+(N_{z}-N_{y})\text{sin}^{2}\phi_{j}\right]\right\}}$ represents the DW width with $A$ the exchange stiffness, $K_{u}$ the magnetocrystalline anisotropy constant, and $\mu_{0}$ being the vacuum permeability. Since we are interested in the genuine oscillation of DW near the pinning notch, we have assumed that the spin torque is absent in Eq. \eqref{Eq2}, while its response to this stimuli can be straightforwardly analyzed (see below).

From the energy point of view, $H_{\text{pin},j}$ can be expressed as the spatial derivative of the total potential:
\begin{equation}\label{Eq3}
H_{\text{pin},j}=-\frac{1}{2\mu_{0}M_{s}L_{y}L_{z}}\frac{\partial U}{\partial q_{j}},
\end{equation}
\begin{figure}[ptbh]
\begin{centering}
\includegraphics[width=0.48\textwidth]{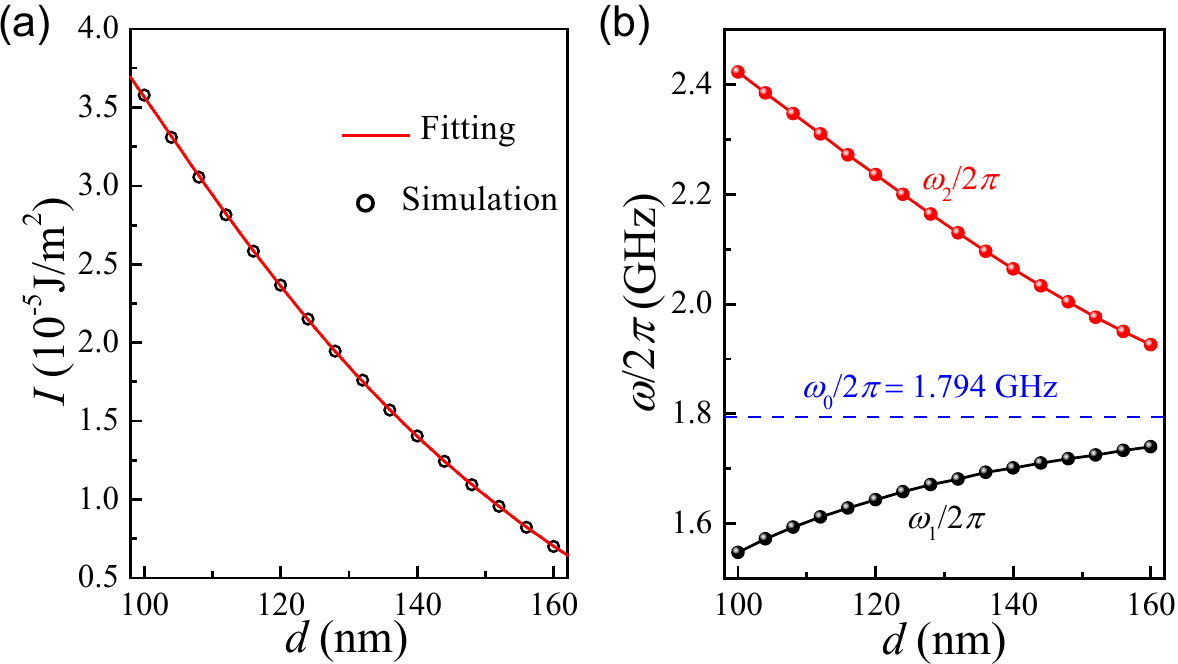}
\par\end{centering}
\caption{(a) Dependence of the coupling strength $\mathcal{I}$ on $d$. Black circles denote simulation results and red solid line represents the analytical fitting. (b) The two eigen frequencies of a DW-DW pair varying with $d$.}
\label{Figure2}
\end{figure}where $L_{y}$ and $L_{z}$ are the width and thickness of the nanostrip, respectively, and $U$ is the total energy of the system including both the pinning potential due to the notches and the coupling between DWs: $U=\sum_{j}\mathcal {K}q_{j}^{2}/2+\sum_{j\neq k}\mathcal {I}(d_{jk})q_{j}q_{k}/2$. Here $\mathcal {K}$ is the spring constant determined by the shape of the notch and $\mathcal {I}(d_{jk})$ is the coupling constant depending on the distance $d_{jk}$ between DWs. Generally, the DW-DW interaction energy can be divided into three parts: the monopole-monopole ($\propto1/d_{jk}$), the exchange ($\propto1/d_{jk}^{2}$), and the dipole-dipole ($\propto1/d_{jk}^{3}$) \cite{PivanoPRB2020}. The explicit form of $\mathcal{I}(d)$ can be obtained from micromagnetic simulations in a self-consistent manner. Considering a small $\phi$ and neglecting the dissipation terms, we arrive at the linear form of Eq. \eqref{Eq2}:
\begin{equation}\label{Eq4}
\mathcal{M}\frac{d^{2}q_{j}}{dt^{2}}+\mathcal {K}q_{j}+\sum_{k\in\langle j\rangle}\mathcal{I}(d_{jk})q_{k}=0,
\end{equation}
where the $\mathcal{M}=2\mu_{0}L_{y}L_{z}/\gamma^{2}(N_{z}-N_{y})\Delta$ is the effective mass of a single DW with $\Delta=\sqrt{2A/\left[2K_{u}+\mu_{0}M_{s}^{2}(N_{y}-N_{x})\right]}$, and $\langle j\rangle$ is the set of the nearest neighbors of $j$. We thus have $\mathcal{I}(d_{jk})=\mathcal {I}_{1}$ ($\mathcal {I}_{2}$) when $j$ and $k$ share an intracellular (intercellular) connection with $\mathcal {I}_{1,2}=\mathcal {I}(d_{1,2})$. Here, $d_{1}$ and $d_{2}$ are alternating intersite lengths. The governing equation is thus mapped to a Su-Schrieffer-Heeger problem \cite{SuPRL1979}. To obtain the analytical formula of $\mathcal{I}(d)$, we simulate the dynamic of a DW-DW pair separated by an arbitrary distance (see Supplemental Material \cite{Suppl} for details). Symbols in Fig. \ref{Figure2}(a) are numerical results and the solid curve is the best fitting $\mathcal{I}(d)=c_{1}/d+c_{2}/d^{2}+c_{3}/d^{3}$, with $c_{1}=-9.2635\times10^{-12}$ J m$^{-1}$, $c_{2}=2.294\times10^{-18}$ J, and $c_{3}=-1.0111\times10^{-25}$ J m. Figure \ref{Figure2}(b) plots the $d$-dependence of the out-of-phase and in-phase DW-oscillation frequencies, that is, $\omega_{1}$ and $\omega_{2}$ respectively, in the simple two-DW system. It shows that $\omega_{1}$ increases while $\omega_{2}$ decreases for an increasing $d$. One naturally expects that $\omega_{1}=\omega_{2}=\omega_{0}$ when $d\rightarrow\infty$, with $\omega_0=\sqrt{\mathcal{K}/\mathcal{M}}$ corresponding to the oscillation frequency of an isolated DW. By measuring $\omega_{0}$ in experiments, one can determine the pinning-potential stiffness $\mathcal{K}$. This approach, however, suffers from an issue that the dynamics of a single DW can be easily modified by structure defects and material randomness, and it thus cannot precisely determine the genuine profile of the pinning potential.
\begin{figure}[ptbh]
\begin{centering}
\includegraphics[width=0.48\textwidth]{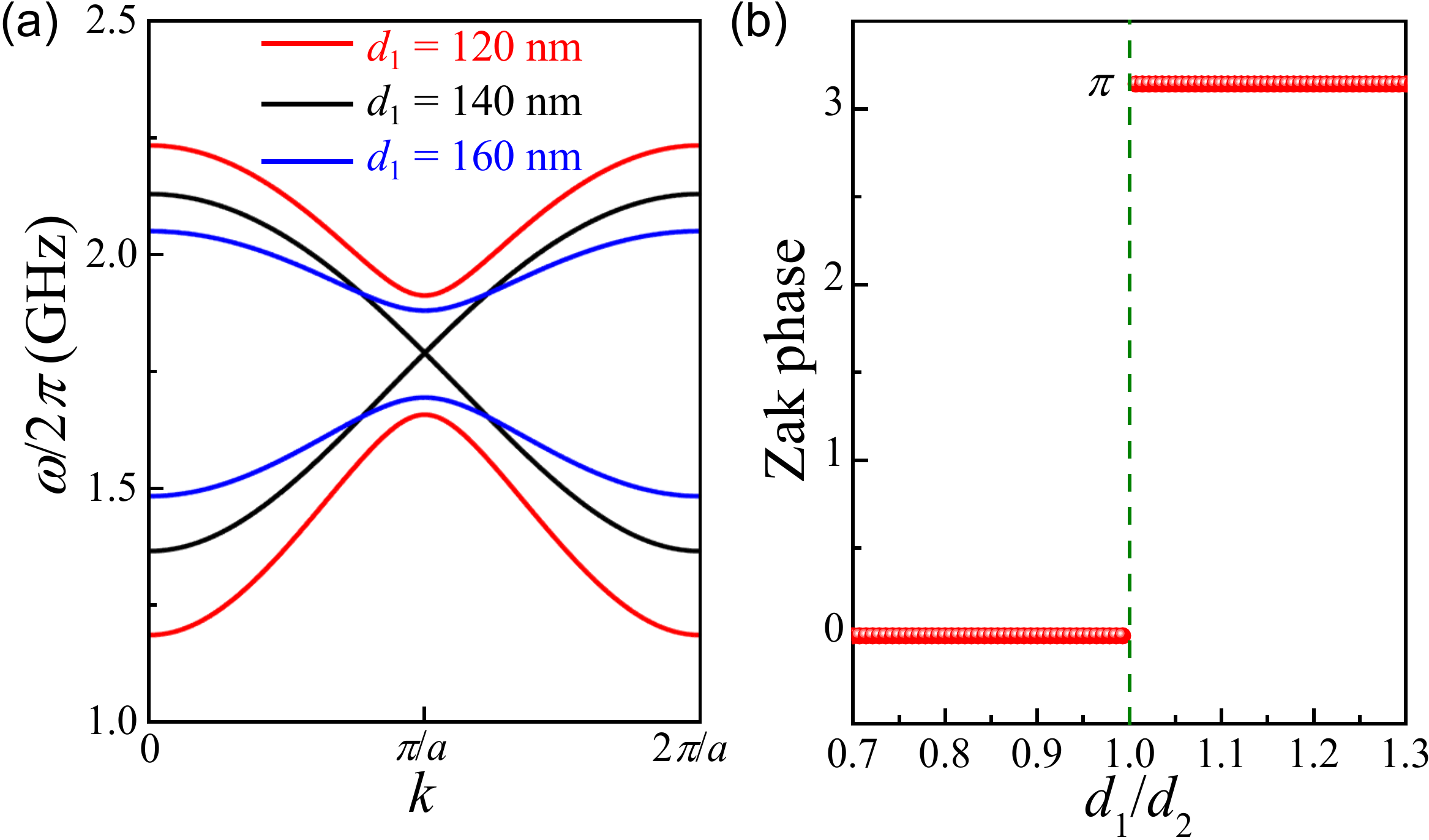}
\par\end{centering}
\caption{(a) Band structure of an infinite DW racetrack for different intracellular lengths: $d_{1}=$120, 140, and 160 nm, with $d_{2}$ being fixed to 140 nm. (b) Dependence of the Zak phase on the ratio $d_{1}/d_{2}$.}
\label{Figure3}
\end{figure}

\emph{Topological phase in DW racetrack.}---For the one-dimensional DW lattice plotted in Fig. \ref{Figure1}(a), the dashed black rectangle represents the unit cell, and the basis vector is $\textbf{a}=a\hat{x}$ with $a=d_{1}+d_{2}$. The band structure of the collective DW oscillations can be computed by a plane wave expansion $q_{j}=q_{j}\exp\big[i(\omega t+nka)\big]$, where $j=A, B$ for different sublattices, $n$ is an integer, and $k$ is the wave vector. The Hamiltonian then can be expressed in momentum space as:
\begin{equation}\label{Eq5}
 \mathcal {H}=\left(
 \begin{matrix}
   \mathcal{K} & \mathcal{I}_{1}+\mathcal{I}_{2}e^{-ika} \\
   \mathcal{I}_{1}+\mathcal{I}_{2}e^{ika} & \mathcal{K}
  \end{matrix}
  \right).
\end{equation}
Solving \eqref{Eq5} gives the dispersion relation:
\begin{equation}\label{Eq6}
\omega_{\pm}(k)=\sqrt{\frac{\mathcal {K}\pm\sqrt{\mathcal {I}_{1}^{2}+\mathcal {I}_{2}^{2}+2\mathcal {I}_{1}\mathcal{I}_{2}\text{cos}ka}}{\mathcal {M}}},
\end{equation}
where $+(-)$ represents the optical (acoustic) branch. The bulk band structures for different geometric parameters are plotted in Fig. \ref{Figure3}(a), where $d_{2}$ is fixed to $140$ nm if not stated otherwise and magnetic parameters of Ni are adopted \cite{Suppl}. For $d_{1}=d_{2}$, the two bands merge together [black curve in Fig. \ref{Figure3}(a)], while a gap opens at $k=\pi/a$ when $d_{1}\neq d_{2}$ [red and blue curves in Fig. \ref{Figure3}(a)], leading to an insulating phase. To judge if these insulating phases are topological, we consider the Zak phase \cite{ZakPRL1985}, a topological invariant that can be evaluated by integrating the Berry connection over the first Brillouin zone:
\begin{equation}\label{Eq7}
\mathbb{Z}=i\int_{0}^{2\pi/a}\Psi^{\dag}(k)\nabla_{k}\Psi(k)dk\ \  (\text{mod}\ 2\pi),
\end{equation}
\begin{figure}[ptbh]
\begin{centering}
\includegraphics[width=0.48\textwidth]{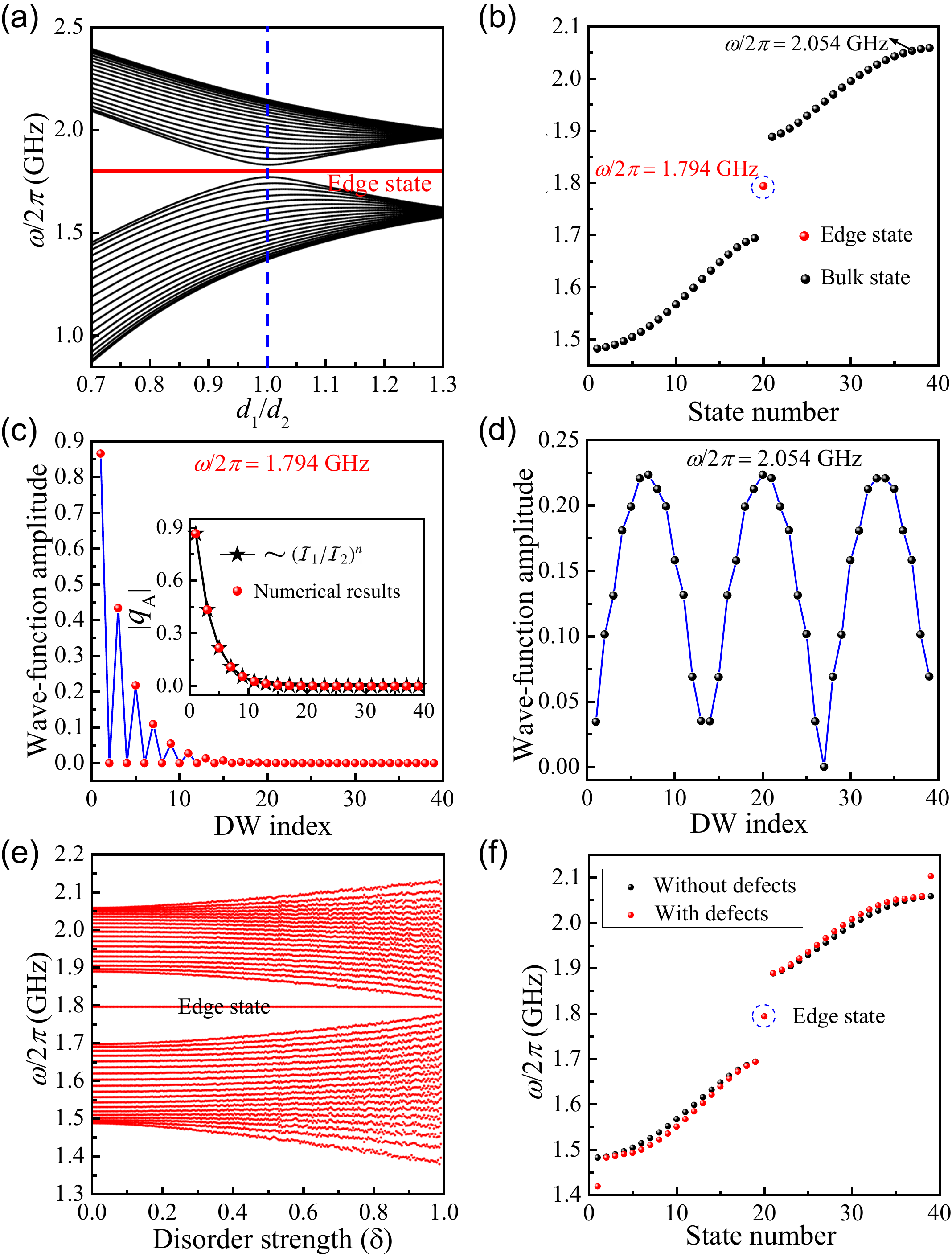}
\par\end{centering}
\caption{(a) Spectrum of a finite DW racetrack for different $d_{1}/d_{2}$. The dotted blue line denotes the boundary separating two topologically distinct phases; the red segment represents the in-gap mode. (b) Eigenfrequencies of the DW lattice with $d_{1}/d_{2}=8/7$. The spatial distribution of DW-oscillation amplitude for the edge (c) and bulk (d) states. (Inset) Comparison between analytical and numerical results. (e) Spectrum of disordered DW racetracks. (f) Spectrum with (red dots) and without (black dots) defects.}
\label{Figure4}
\end{figure}where $\Psi(k)$ is the Bloch wave function of the energy band. Figure \ref{Figure3}(b) shows the dependence of the Zak phase $\mathbb{Z}$ on the ratio $d_{1}/d_{2}$. It is observed that $\mathbb{Z}$ is quantized to 0 when $d_{1}/d_{2}<1$ and to $\pi$ otherwise, indicating two topologically distinct phases in the two regions. We point out that this conclusion is independent of the choice of $d_{2}$.

Bulk-boundary correspondence indicates the existence of robust edge states. To verify this point, we compute the spectrum of a finite racetrack containing an odd number (e.g., 39) of DWs. Numerical results are shown in Fig. \ref{Figure4}(a), where the in-gap state (red line) emerges for all ratios $d_{1}/d_{2}\neq1$. We first consider the case $d_{1}/d_{2}=8/7\ (>1)$. Figure \ref{Figure4}(b) plots the eigenfrequencies of the system, showing that there is one in-gap mode marked by red dot. Further, it is found that its spatial distribution is highly localized at the left end of the racetrack [see Fig. \ref{Figure4}(c)], in contrast to its bulk counterpart shown in Fig. \ref{Figure4}(d). We adopt the \emph{Ans\"{a}tze} for the localized mode as $q_{j}=q_{j}\text{exp}(i\omega_{0} t)z^{n}$ with $|z|<1$. The edge state then can be solved by the equations:
\begin{equation}\label{Eq9}
\begin{aligned}
(\mathcal{I}_{1}+\mathcal{I}_{2}z)q_{A}(n)&=0,\ \text{for}\  n=2,3,...\\
(\mathcal{I}_{1}+\mathcal{I}_{2}z^{-1})q_{B}(n)&=0,\ \text{for}\  n=1,2,3,...
\end{aligned}
\end{equation}
\begin{figure}[ptbh]
\begin{centering}
\includegraphics[width=0.48\textwidth]{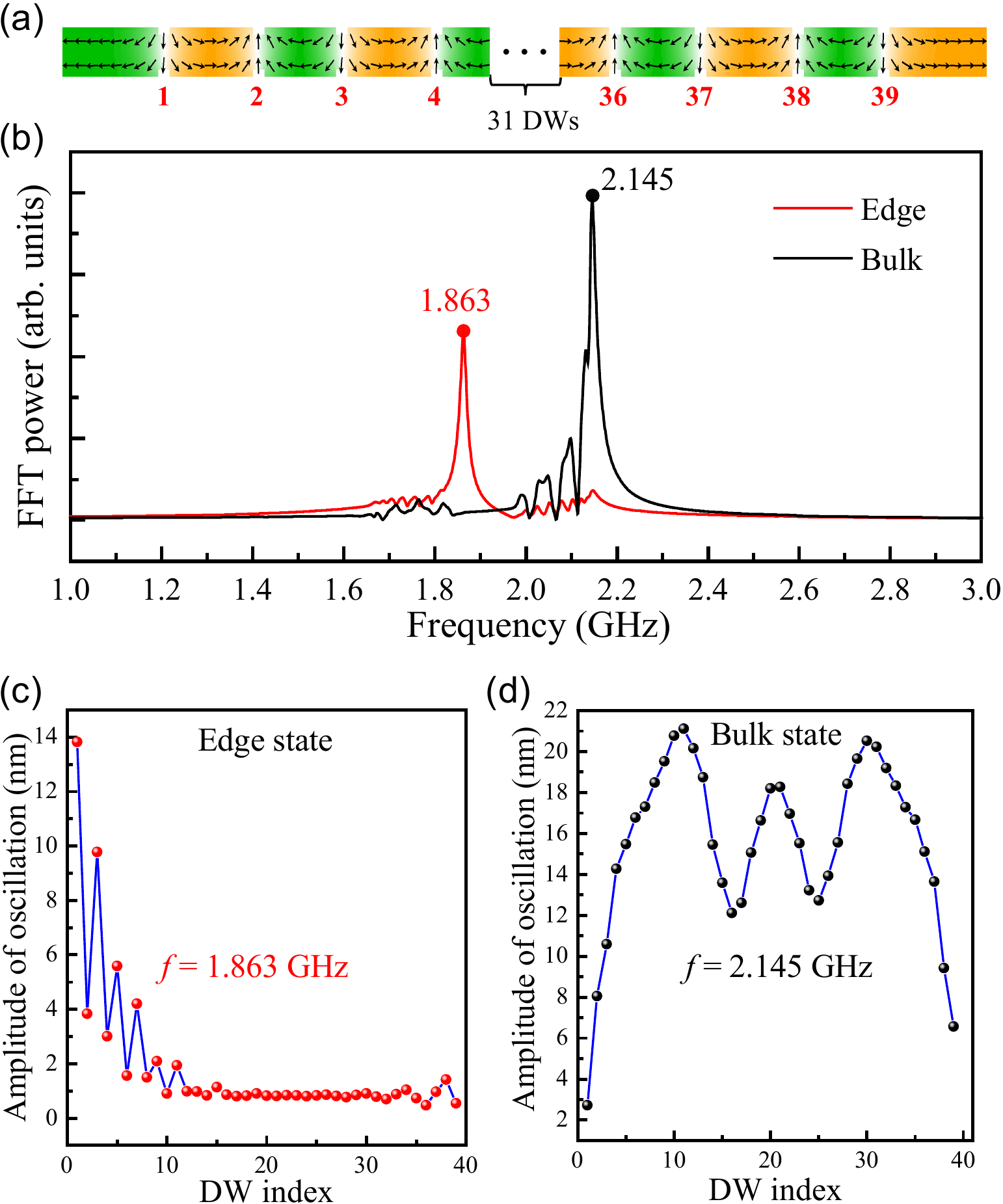}
\par\end{centering}
\caption{(a) Schematic plot of a finite racetrack containing $39$ DWs, with $d_{1}=160$ nm and $d_{2}=140$ nm. (b) The temporal Fourier specra of the DWs oscillation at edge (1st DW) and bulk (20th DW) positions. The spatial distribution of amplitude of DW oscillations for edge (c) and bulk (d) states.}
\label{Figure5}
\end{figure}with the boundary condition $\mathcal{I}_{1}q_{B}(1)=0$ (A-site DW is in the outmost left boundary).
Because $\mathcal{I}_{1,2}\neq 0$, we obtain $q_{B}(n)=0$ $\forall n$ and $z=-\mathcal{I}_{1}/\mathcal{I}_{2}$. The wave-function amplitude of A-site DWs therefore follows an exponentially decaying formula $|q_{A}|=|q_{0}|(\mathcal{I}_{1}/\mathcal{I}_{2})^{n}$ for $n=1,2,3,...$. Analytical result agrees excellently with numerical calculations, as plotted in the inset of Fig. \ref{Figure4}(c). We point out that the edge state becomes localized in the right end instead if $d_{1}/d_{2}<1$ (not shown). Interestingly, the localized modes emerge in both ends as the magnetic racetrack contains an even number of DWs (detailed calculations can be found in Supplemental Material \cite{Suppl}).

To verify the topological robustness of the edge states, we calculate the spectrum of the DW racetrack including disorder and defects, with results presented in Figs. \ref{Figure4}(e) and \ref{Figure4}(f), respectively. Here the disorder is introduced by assuming that the coupling parameters $\mathcal{I}_{1}$ and $\mathcal{I}_{2}$ have a random variation, i.e., $\mathcal{I}_{1}\rightarrow \mathcal{I}_{1}(1+\delta N)$, $\mathcal{I}_{2}\rightarrow \mathcal{I}_{2}(1+\delta N)$, with $\delta$ the disorder strength and $N$ a uniformly distributed random number between $-1$ and 1. As to the defects, we assume $\mathcal{I}_{1}$ and $\mathcal{I}_{2}$ suffering from a shift ($\mathcal{I}_{1}\rightarrow 10\mathcal{I}_{1}$, $\mathcal{I}_{2}\rightarrow 0.1\mathcal{I}_{2}$) on the second and fourth DWs. From Figs. \ref{Figure4}(e) and \ref{Figure4}(f), we observe that the edge state is very robust against these disorder and defects, while the bulk states are sensitive to them.

\emph{Micromagnetic simulations.}---To confirm our theoretical predictions, we use the micromagnetic package MUMAX3 \cite{VansteenkisteAdv2014} to simulate the dynamics of $39$ interacting DWs in Ni nanostrip of length $7000$ nm, as shown in Fig. \ref{Figure5}(a) \cite{Suppl}. To obtain the spectra of DW oscillations, a sinc-function magnetic field $H(t)=H_{0}\sin[2\pi f_{0}(t-t_{0})]/[2\pi f_{0}(t-t_{0})]$ is applied for 1 $\mu$s along the $x$-axis with $H_{0}=10$ mT, $f_0=20$ GHz, and $t_{0}=1$ ns. The position of all DWs ${q}_{j}$ is recorded every 100 ps, where we define $q_{j}=\int \!{x\,|m_{y}|^{2}dx}/\int \!{|m_{y}|^{2}dx}$ with the integration confined at the $j$-th DW. Here, $|m_{y}|$ is adopted as the weight function based on the fact that the closer to the center of DW, the greater the $y$ component of the magnetization.

To find the frequency range of the edge and bulk states, we analyze the temporal Fourier spectra of the DW racetrack at two different positions (DW 1 and DW 20, for example). Figure \ref{Figure5}(b) shows the results, with peaks of the red and black curves denoting the positions of edge and bulk bands, respectively. We then apply a sinusoidal magnetic field $\textbf{h}(t)=h_0\sin(2\pi ft)\hat{x}$ with $h_{0}=0.05$ mT over the whole system to excite the edge and bulk modes by choosing two frequencies $f=1.863$ GHz and $f=2.145$ GHz, respectively, as marked in Fig. \ref{Figure5}(b). The spatial distribution of DW oscillation amplitude for these two modes are plotted in Figs. \ref{Figure5}(c) and \ref{Figure5}(d), respectively, from which one can clearly identify the localized and extended nature of edge and bulk states, respectively. Full micromagnetic simulations are thus well consistent with the analytical results but with the following discrepancies: (i) The oscillation amplitude of B-site DWs does not exactly vanishes as dictated by the analytical theory, but is promoted by A-site DW oscillations. We attribute it to the anharmonic DW-DW interaction \cite{Suppl}; (ii) A finite DW oscillation is observed in the deep bulk of the racetrack, as opposed to an exponential decay to zero. However, we note that the DW oscillation amplitude in such case is smaller than the mesh size (2 nm) adopted in the micromagnetic simulations. It is thus still within the numerical accuracy.

\emph{Discussion and conclusion.}---By including the STT term in Eq. \eqref{Eq2}, we find that it does not change the spectrum of the collective DW oscillations, but causes a global shift $X=\beta\gamma(N_{y}-N_{z})M_{s}b_{J}/\omega^{2}_{0}$ to the equilibrium DW position \cite{Suppl}. It is noted that the imaging of the DW position is already within current technology reach. Since the DW oscillation frequency $\omega_{0}$ can be standardized by the topological method, we are able to accurately quantify the STT non-adiabaticity coefficient $\beta$ by experimentally measuring the slope of $X-b_{J}$ curve.

In summary, we studied the Su-Schrieffer-Heeger problem in a one-dimensional DW racetrack with periodic pinning notches. Zak phase was evaluated to derive the phase diagram that allows two topologically distinct phases separated by the phase transition point at an identical intercellular and intracellular length between neighbouring DWs. The emerging edge state dictated by the bulk-boundary correspondence was shown to be particularly robust against moderate defects and disorder. Analytical results were well supported by full micromagnetic simulations. We propose that the uncovered topological feature can be utilized as the DW frequency standard, which shall encourage our experimental colleagues to accurately measure the pinning profile and to finally resolve the controversy about the $\beta$ parameter. Our findings suggest as well that the DW racetrack offers an ideal platform to explore the fundamental topological phase, such as the higher-order topological corner (hinge) states in two-(three-)dimensional structures, and the nonlinear effect in topological phase transition, which are interesting issues for future study.

\begin{acknowledgments}
This work was supported by the National Natural Science Foundation of China (NSFC) (Grants No. 12074057, No. 11604041, and No. 11704060). Z.-X. Li acknowledges financial support from the China Postdoctoral Science Foundation (Grant No. 2019M663461) and the NSFC (Grant No. 11904048). Z. Wang was supported by the China Postdoctoral Science Foundation under Grant No. 2019M653063.
\end{acknowledgments}

\end{document}